\documentclass[a4paper, prl, twocolumn, showpacs]{revtex4}
\usepackage{float}
\restylefloat{table}

\usepackage{amssymb}
\usepackage{amsmath}
\usepackage{color, mathrsfs, amsmath,amscd}
\newcommand{\hilight}[1]{\colorbox{yellow}{#1}}
\usepackage{graphics, graphicx}
\usepackage{bbold}

\usepackage[normalem]{ulem}
\newcommand\redout{\bgroup\markoverwith
{\textcolor{red}{\rule[.5ex]{2pt}{0.4pt}}}\ULon}
\usepackage{xcolor}

\usepackage{verbatim} 

\newcommand{\mr}{\mathrm}

\newcommand{\bb}{\mathbb}

\usepackage{psfrag}
\usepackage{mathcomp}
\usepackage{subfigure}

\usepackage{color}
\usepackage{soul}

\begin{document}

\author{Guowu Meng}

\affiliation{Department of Mathematics, HKUST, Clearwater Bay, Kowloon, Hong Kong} 
\date{April 11, 2013}

\thanks{The author was supported by the Hong Hong Research Grants Council under RGC Project No. 603110 and the Hong Kong University of Science and Technology under DAG S09/10.SC02.}

\title{5th Force and Quark Mixing}

\begin{abstract}
In a recent article, this author proposed a program for physics beyond the Standard Model, solely based on modifying the twin pillars of fundamental physics by replacing Lorentz structure with Euclidean Jordan algebra while keeping quantum theory. This program predicts not only quarks and leptons but also a short-range 5th fundamental force accompanying gravity.

This 5th force predicts quark mixing and the related CP violation, which in fact was a phenomena observed in labs about fifty years ago. Thus, there are two conflicting theories as of now, the one based on the 5th force which {\em predicts} this phenomena and the established Cabibbo-Kobayashi-Maskawa (CKM) theory which was invented to {\em explain} this phenomena. In this article a test of these two theories against the recent experimental data is presented. It is found in this test that the CKM theory fares poorly, whereas the one based on the 5th force withstands the test well, in both accuracy and precision.   For example, for the CKM matrix entry $V_{cb}$, we have
$$|V_{cb}^{\tiny \mbox{experiment}}|=0.0409 \pm   0.0011, \quad |V_{cb}^{\tiny \mbox{CKM}}|= 2.37\pm 1.82,\quad |V_{cb}^{\tiny \mbox{5th force}}|=0.0408 \pm   0.0028.
$$

 \end{abstract}
 
\maketitle

\section{introduction}
It was experimentally observed in the 1960s that a quark of high generation can decay into a quark of low generation: 
\begin{center}
{
\includegraphics[scale=0.65]{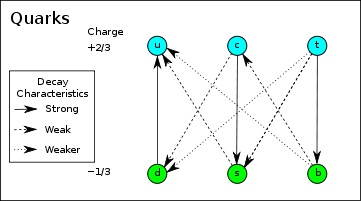}
A diagram of the mixed quark decay. (http://en.wikipedia.org/wiki/User:Army1987/Quark)
}
\end{center}
\vskip 5pt
This phenomenon,  referred to as {\it the mixed quark decay} in this article, has been thought to be explained by the CKM theory \cite{Cabibbo63, KM73} in which it is assumed that the weak interaction pairs are $(u, d')$, $(c, s')$ and $(t,b')$ rather than the more natural pairs $(u, d)$, $(c, s)$ and $(t,b)$, where
$$
\left[\begin{matrix}
d'\cr s' \cr b'
\end{matrix}\right]=V\left[\begin{matrix}
d\cr s \cr b
\end{matrix}\right] 
$$
with $V$ being the CKM matrix
$$
V=\left[\begin{matrix}
V_{ud} & V_{us} & V_{ub}\cr
V_{cd} & V_{cs} & V_{cb}\cr
V_{td} & V_{ts} & V_{tb}
\end{matrix}\right].
$$
For example, via weak interaction, instead of decaying into $b$, $t$ decays into $b'= V_{td}d+V_{ts}s+V_{tb}b$, and hence decays into $d$, $s$ or $b$ with the relative probability $|V_{td}|^2$, $|V_{ts}|^2$ and $|V_{tb}|^2$ respectively. In this theory, the CKM matrix $V$ is simply an input parameter whose actual value must be determined by experiments, but one thing is clear: $V$ has to be unitary \emph{exactly}. A great success of this theory is the prediction of the third generation of quarks, for which Kobayashi and Maskawa shared one half of the 2008 Nobel Physics prize in Physic \cite{NobelPrize08}.

From the particle data group \cite{PDG2012}, the best experimental determination for the magnitudes of the CKM matrix elements is $V^{exp}=$
  \begin{eqnarray}\label{experiment}
{\small \left[\begin{matrix}
0.97425\pm 0.00022 & 0.2252\pm 0.0009 &0.00415\pm 0.00049\cr
0.230\pm 0.011 & 1.006\pm 0.023 & 0.0409\pm 0.0011\cr
\hilight{$0.0084\pm 0.0006$} & \hilight{$0.0429\pm 0.0026$}  & 0.89\pm 0.07
\end{matrix}\right]. }
 \end{eqnarray}
It is worth to mention that, the values of $|V_{td}|$ and $|V_{ts}|$ quoted here (the highlighted matrix entries) are obtained under the theoretical assumption that $|V_{tb}|=1$, see equation (11.13) of page 159 in Ref. \cite{PDG2012} and the line above this equation. If we took the experimentally determined value of $|V_{tb}|$, i.e., $0.89\pm 0.07$, we would have
 \begin{eqnarray}\label{rescaled}	
 |V_{td}| =0.0075\pm 0.0008, \quad |V_{ts}|= 0.0382\pm 0. 0038.
 \end{eqnarray}

The phases of the CKM matrix elements determine the amount of CP violation. With the observed amount of CP violation taken into account, the currently best known ``standard" fit for the CKM matrix elements in the sense of Chau and Keung \cite{Chau84} is $V^{fit}\approx$
\begin{eqnarray}
\tiny \left[\begin{matrix}
0.9742 & 0.2256 & \hilight{$0.00127 - 0.00326i$}\cr
-0.2255-0.0001i & 0.97336-0.00003i& 0.04153\cr
\hilight{$0.00814-0.00317i$} & -0.0407-0.0007i & 0.9991 \cr
\end{matrix}\right]. \nonumber
\end{eqnarray}
Note that only the two highlighted entries are not nearly being real numbers. Because of this observation, \emph{$V_{ud}$, $V_{us}$, $V_{cd}$, $V_{cs}$, $V_{ts}$  shall be assumed to be real numbers hereafter}.

The assumption that $t$ interacts with $b'$ (rather than $b$) via weak interaction is made purely for explaining the mixed quark decay. There is no theoretical reason why this must be so. In any case, this assumption is mysterious and now the experimental data strongly suggests that it is incorrect. For example, the norm square of the 3rd column should be exactly $1$; however, based on the experimental data for the 3rd column, a simulation of any size $> 10^6$ shows that this norm square has an average $\approx 0.80$, a standard deviation $\approx 0.12$, and is less than $0.9$ more than $79\%$ of time.

It is perhaps not a secret among some experts that the CKM theory is less than accurate. However,
to reject the CKM theory, one must provide an alternative theory which is both theoretically sound as well as quantitatively more accurate. A major message in this article is that such a theory indeed exists.

\section{5th Force Based Theory}
In a recent article \cite{meng12} of this author, it was shown that one can indeed derive the existence of quarks and leptons provided that one is willing to make a new modification for the twin pillars of fundamental physics, as shown in the last stage of the following chain of modifications:
\begin{center}
 \redout{\em Classical mechanics} + {\em Galilean structure}
$$\downarrow\begin{matrix} \mbox{quantum}\\ \mbox{modification}\end{matrix}$$
{\em Quantum theory} +  \redout{\em Galilean structure}
$$\downarrow \begin{matrix} \mbox{relativity}\\ \mbox{modification}\end{matrix}$$
{\em Quantum theory} + \redout{\em Lorentz structure}
$$\downarrow \begin{matrix} \mbox{new}\\ \mbox{modification}\end{matrix}$$
{\em Quantum theory} + {\em Euclidean Jordan algebra}.
\end{center}
This new modification is comparatively conservative because Lorentz structure is a secondary structure hidden inside Euclidean Jordan algebra, rather than an approximation to it. With this new modification, quarks and leptons, the four fundamental forces and the broken electric-weak symmetry, matter generations, and other experimentally found phenomena, appear naturally in the theoretical framework. Moreover, a short-range 5th fundamental force accompanying gravity is predicted.

This 5th force is predicted to violate the CP symmetry and transform quarks among its various generations, so it immediately predicts mixed quark decay and the related CP violation. For example, since $s$ decays to $d$ via the 5th force and $d$ decays to $u$ via the weak force, $s$ must decay to $u$. 

\section{Quantitative Checks with Experiments}
The theoretical framework offered above sounds very simple and leaves no mysteries behind, but in order for this to be a valid theory it must be able to withstand quantitative checks with experiments. Since the precise mathematical form for the 5th force is not available at the moment, one has to make some reasonable assumptions on the 5th force in order to proceed. With this in mind, by using simple rules in quantum mechanics one can derive the following \emph{inexact} formulae for $V_{tb}$,  $V_{cb}$, $V_{td}$:
\begin{eqnarray}\label{formula}
\hilight{\fbox{$
\begin{array}{rcl}
V_{tb} &= & 3V_{ud}-2V_{cs},\\
\\
V_{cb} &= & {V_{cs}V_{us}-V_{cd}V_{ud}\over V_{us} (V_{cs}+V_{tb})} V_{ts},\\
\\
V_{td} &=& {V_{cs} V_{us}V_{ts}\over (V_{cs}+V_{tb})V_{ud}}+{V_{ud}V_{tb} V_{ub}^*-V_{us} V_{cb}V_{tb}\over V_{ud}^2}+w.
\end{array}$}}
\end{eqnarray} Here, $*$ means the complex conjugate, and $w$ is some unknown complex constant. To test the accuracy of these formulae, one takes as input data the experimentally determined magnitude for the matrix entries entered into the right hand side of formulae (\ref{formula}), but with this sign convention chosen: diagonal entries are all positive, $V_{us}>0$ and $V_{ts}<0$. To be more precise, we take
$V^{exp}_{incomplete}=$
\begin{eqnarray}\label{experiment}
{\tiny \left[\begin{matrix}
0.97425\pm 0.00022 & 0.2252\pm 0.0009 & (0.00415\pm 0.00049) e^{-i(1.2\pm 0.08)} \cr
\\
-0.230\pm 0.011 & 1.006\pm 0.023 & \cr
\\
 &   &  
\end{matrix}\right] }\nonumber
 \end{eqnarray}
and ${V_{ts}\over |V_{tb}|}=-0.0429\pm 0.0026$ (page 159 in Ref. \cite{PDG2012}) as our input data, then compute $V_{tb}$, $V_{cb}$ and $V_{td}$ according to formulae (\ref{formula}). Note that $V_{td}$ cannot be computed from the formulae because $w$ is unknown, however, its standard deviation, which is independent of the actual value of $w$, can be computed. In our actual computation we set $w=0$. 

In the CKM theory, the orthogonality for the first two columns and first two rows plus the unity of the norm square of the 3rd row yields the following very accurate formulae for $V_{ub}$, $V_{cd}$ and $V_{tb}$:
\begin{eqnarray}\label{CKMformula}
\begin{array}{rcl}
|V_{cb}| &=& {|V_{ud}V_{cd}+V_{us}V_{cs}|\over |V_{ub}|},\\
\\
|V_{td}| &=& \sqrt{|V_{ud}V_{us}+V_{cd}V_{cs}| {|V_{td}|\over |V_{ts}|}}\\
\\
|V_{tb}| &= & \sqrt {1- |V_{td}|^2(1+{|V_{ts}|^2\over |V_{td}|^2}) }.
\end{array}
\end{eqnarray}
As the input data for the computation via formulae (\ref{CKMformula}), we take
$V^{exp}_{incomplete}$ as before and ${|V_{td}|\over |V_{ts}|}=0.221\pm 0.07$, see page 159 in Ref. \cite{PDG2012}.

By performing a numerical test with a sample size $ > 10^6$, one gets the following result: 
\begin{table}[H]
    \begin{tabular}{|c|c|c|c|}
        \hline
       & {$\begin{matrix}\mbox{CKM prediction}  \cr
       \mbox{per formulae (\ref{CKMformula})}\end{matrix}$} &  {$\begin{matrix} \mbox{5th force prediction} \\
       \mbox{per formulae (\ref{formula})}\end{matrix} $}  & \mbox{experiment}\\ \hline
        $|V_{tb}|$ & $0. 9657 \pm 0.0244$  &  $ 0.911\pm 0.046$ & $0.89\pm 0.07 $\\ 
        \hline
        $|V_{cb}|$ & $2.37\pm 1.82$ &  $ 0.0408\pm 0.0028$ & $0.0409\pm 0.0011 $\\ 
        \hline

        $|V_{td}|$& $0.05\pm0.02$ & $\hilight{$0.0048$}\pm 0.0005$ &  $0.0084\pm 0.0006$ \\ 
           \hline      
    \end{tabular}
\end{table}
\vskip -15pt
\noindent Note that the uncertainties in the output come from the uncertainties of the experimental data for the matrix entries entered into the right hand side of formulae (\ref{formula}) and (\ref{CKMformula}). In the tests,  each CKM matrix entry on the right hand side of formulae (\ref{formula}) or  formulae (\ref{CKMformula}) is assumed to be a Gaussian random variable with the normal distribution determined by the experimental data. For example, $V_{ud}$ is assumed to be the Gaussian random variable with the normal distribution having average equal to $0.97425$ and standard deviation equal to $0.00022$.

In the 5th force based predictions, the precision is good, so is the accuracy except for $|V_{td}|$. The inaccuracy for $|V_{td}|$ has a known reason:  the unknown complex constant $w$ in formulae (\ref{formula}) was set to be zero in the numerical test. In contrast, in the CKM based predictions, neither precision nor accuracy fares well, especially for $|V_{cb}|$ and $V_{td}|$.  

In summary, the \emph{inaccurate} formulae deduced from the 5th force produces a much better result than the CKM theory's \emph{accurate} formulae. Thus, considering this result, as well as observation that the 5th force theory is natural and is also part of a bigger natural theory proposed in Ref. \cite{meng12}, the 5th force theory does a better job than the CKM theory at predicting mixed quark decay.

\section{Derivation of Formulae (\ref{formula})}
The goal of this section is to derive formulae (\ref{formula}) based on the 5th force and the weak force. Here the 5th force transforms UP quarks (DOWN quarks respectively) among their different generations, but the weak force is still the one in the classical Standard Model, i.e., it transforms UP quarks and DOWN quarks into each other only in the same generation. In contrast, in the CKM theory, since the mixed quark decays is solely due to the weak force, the weak force must be assumed to transform a UP quark into a linear supoposition of DOWN quarks from various generations. Apparently these are two conflicting theories for mixed quark decays.

For the purpose in this article,  we are only interested in computing the following transition amplitude matrix:
$$
A:=\left[\begin{matrix}
A_{du} & A_{su} & A_{bu}\cr
A_{cd} & A_{cs} & A_{bc}\cr
A_{td} & A_{ts} & A_{tb}
\end{matrix}\right]. 
$$
Here $A_{ab}$ denotes the total transition amplitude (due to both the weak force and the 5th force) for quark $a$ spontaneously decaying into quark $b$.  Note that the matrix $A$ is a scalar multiple of $V$ up to the complex conjugation of some entries. For the purpose of comparing with $V$,  hereafter we shall assume that $A=V$ modulo the complex conjugation of some entries.

The precise form of the 5th force is not known at the moment, otherwise one would be able to compute $A$ directly. In spite of this fact, based on rules in quantum theory and some plausible physics assumptions, one can still derive some useful information. 

As a simple exercise, let us derive the following formula 
\begin{eqnarray}\label{simpleF}
A_{tb} = 3A_{du}-2A_{cs}
\end{eqnarray}
which corresponds to the simplest identity in formulae (\ref{formula}). It is reasonable to assume that the weak force transition amplitude between the UP quark and the DOWN quark in each generation is a same positive number $A_w$ when suitable relative phases are chosen. Then
$$
A_{du} =A_w, \; A_{cs}= A_w+A',\; A_{tb} = A_w+ A''+ A'''.
$$
Here $A'$ is the transition amplitude contributed from the decaying process involving $c$ first decaying to $u$ via the 5th force,  $A''$ is the transition amplitude contributed from the decaying process involving $t$ first decaying to $u$ via the 5th force, and $A'''$ is the transition amplitude contributed from the decaying process involving $t$ first decaying to $c$ via the 5th force. It is reasonable to assume that  $A', A'', A'''$ have the same amplitude and are real, then $A_{tb} = A_w \pm  A' \pm A'$. Among the four possible choices of signs, only this one
$$
A_{tb} = A_w -  A'  - A'
$$
produces result that matches the experiment well, so $A_{tb} = A_w -  2A'$.  Then $2A_{cs}+A_{tb}=3A_{du}$,  which is essentially formula (\ref{simpleF}). 

Note that, because of the sign ambiguity one actually has four formulae for $A_{ts}$, but the one we chose is the one that matches experiment the best.  

To derive the complicated identities in formulae (\ref{formula}), let us introduce a few more notations. Denote by $A_{ab}^f$ the 5th force transition amplitude from quark $a$ to quark $b$. Here $a$ and $b$ are either all UP quarks or all DOWN quarks. Let
\begin{eqnarray}
x_1 = A^f_{tc}, \quad  x_1' = A^f_{cu}, \quad  x_1'' = A^f_{tu}, \cr
x_2 = A^f_{bs}, \quad x_2' = A^f_{sd}, \quad x_2'' = A^f_{bd}, \nonumber
\end{eqnarray}
and
\begin{eqnarray}
y_1 &= &\mbox {the transition amplitude from $d$ to $u$},\cr
y_2 &= &\mbox {the transition amplitude from $c$ to $s$},\cr
y_3 &= &\mbox {the transition amplitude from $t$ to $b$},\cr
z &= &\mbox {the transition amplitude from $b$ to $c$ via $s$},\cr
z' & = &\mbox{the transition amplitude from $c$ to $d$ via $u$},\cr
z''  &= & \mbox{the transition amplitude from $t$ to $d$ via  $u$}.\nonumber
\end{eqnarray}

These are of course the simple transition amplitudes. For the convenience of readers, some of these transition amplitudes have been recorded in the following diagram:
$$\begin{CD}
u @<x_1'<< c @<x_1<<t\\
y_1@AAA y_2@VVV y_3@VVV\\
d @<x_2'<< s@<x_2<< b
\end{CD}
$$

In the next step, we use the basic rules in quantum mechanics to figure out the interesting transition amplitudes in terms of the preceding simple transition amplitudes. Of course, we must ignore gravity, the electric force and the strong force in our analysis here. 

These transition amplitudes are computed according to the rules in quantum mechanics. For example, to compute $A_{ts}$, one notes that there are exactly two routes from $t$ to $s$, one via $c$ which contributes $x_1y_2$ to $A_{ts}$ and one via $b$ which contributes $y_3x_2$ to $A_{ts}$, so 
$$
A_{ts}= x_1y_2+y_3x_2.
$$
As another example,  to compute $A_{bc}$, one notes that there is only one route from $b$ to $c$, i.e., the one via $s$ which contributes $z$ to $A_{bc}$,  so 
$$
A_{bc}= z.
$$
(The route via $t$ is not available because $b$ cannot spontaneously decay to $t$ which is heavier than $b$.)
A few more exercise of this kind yields the following equation
{\scriptsize   \begin{eqnarray}\label{matrix}
A= \left[\begin{array}{ccc}
y_1 & x_2'y_1 &  (x_2x_2'+x_2'')y_1+x_1'z\\
\\
x_2'y_2+z'& y_2  & z\\
\\
y_3(x_2x_2'+x_2'')+\\x_1(y_2x_2'+z')+z'' & x_1y_2+x_2y_3 & y_3
\end{array}\right]
\end{eqnarray}}
The matrix $A$ at the moment is useless as far as predictions are concerned. However, one notes that any match with experiment has to be more than just numerical, it must meet all reasonable physics expectations, too. For example, with suitable relative phases fixed,  it is very reasonable to expect that, up to sign,  $x_1\sim x_2$ (say $x$), $x_1'\sim x_2'$ (say $x'$), and $x_1''\sim x_2''$ (say $x''$), $y_1\sim y_2\sim y_3\sim 1$, ${z'\over z}\sim {x'\over x}$ (say $r_1$). Further, if we assume hydrogen atom's spontaneous transition amplitudes hold roughly, we expect $r_1\sim 10$ and $r_2:={x''\over x}\sim 1$.  This last physics expectation is reasonable in view of Ref. \cite{meng12} in which the elementary particles are proposed to be modeled on the hydrogen atom.  With this in mind, a numerical match with $r_2$ which is either nearly zero or too big would not be considered as a match on physics ground.

In the further analysis, it shall be assumed
\begin{eqnarray}\label{assumption}
\fbox{$
\begin{array}{ccl}
x_1 \sim x_2 \sim x,& x_1' \sim x_2' \sim x', & x_1'' \sim x_2'' \sim x'',\cr
 \quad z'x+zx'\sim 0. & & 
 \end{array}
 $}
\end{eqnarray}
Then 
 {\tiny \begin{eqnarray}
A \sim \left[\begin{array}{ccc}
y_1 & x'y_1 &  (xx'+x'')y_1+x'z\\
\\
x'y_2-z{x'\over x}& y_2  & z\\
\\
x'x(y_2+y_3)+y_3x''+w & x(y_2+y_3) & y_3 
\end{array}\right]\end{eqnarray}}
where $w=z''+xz'$.
Note that, there are seven variables here: $x$, $x'$, $x''$, $z$, $y_1$, $y_2$ and $y_3$, so $A$ has predictive power now, that is because, with the identification of $A$ with $V$, this form of $A$ yields the complicated identities in formulae (\ref{formula}). Note also that, if $V_{td}$ and $V_{cb}$ are both positive, the match with experiments is also good numerically, but then the ratio $x''\over x$ is very tiny, so has to be rejected on physics ground; on the other hand, if $V_{td}$ and $V_{cb}$ are both negative, the match with experiments is good both numerically and on physics ground because ${x''\over x}\approx 0.96$ is indeed close to $1$ in this case. 

\section{Conclusion}
Thanks to the more refined experimental data existing today, the mixed quark decay is better explained by a 5th force based theory than the CKM theory. Since the 5th force is quite weak, its effect is very small so that only in certain special cases such as the one examined here can one possibly find a serious mismatch between the Standard Model and experiments \footnote{CP violations in the decay of Muons is predicted. However, according to the experimental data used in this article, about $5\%$ of the decay of Muons is due to the 5th force. In view of the amount ($\sim {1\over 500}$) of CP violation observed for the decay of Kaons, the actual amount of CP violation in the decay of Muons is about ${1\over 500} \times 5\%$, i.e., about $0.01\%$, which is too small to be observed.}. 

The program proposed in Ref. \cite{meng12} is based on a conservative modification of the twin pillars of fundamental physics. The research presented here provides a successful quantitative test of this program, and in the process serious flaws in the established CKM theory are exposed. 

\section{acknowledgements}
I would like to thank my colleagues K. Chen, B. Y. Jing, J. S. Li, Q. M. Shao, T.Z. Qian, M. Yan, L. J. Zhu for helpful discussions and advices, and Z. Ligeti for a clarification of the fit result for CKM matrix in Ref. \cite{PDG2012}.  I would also like to thank C. Taubes for thoughtful discussions, advices, encouragement, and various important suggestions on improving this article.

\end{document}